\documentclass[letter]{aa}

\usepackage{graphicx}
\usepackage{hyperref}
\hypersetup{colorlinks, allcolors=blue}
\usepackage[varg]{txfonts}

\begin{document}

    \title{From moving groups to star formation in the Solar Neighborhood}
    \titlerunning{From moving groups to star formation}
    \authorrunning{C. Swiggum et al.}

    \author{C. Swiggum\inst{1},
            J. Alves\inst{1},
            E. D'Onghia\inst{2,3}
          }

    \institute{
             Department of Astrophysics, University of Vienna, Türkenschanzstrasse 17, 1180 Wien, Austria\\
                \email{cameren.swiggum@univie.ac.at}
        \and
        Department of Astronomy, University of Wisconsin-Madison, 475 North Charter Street, Madison, WI 53706, USA
        \and
        INAF - Osservatorio Astrofisico di Torino, via Osservatorio 20, 10025 Pino Torinese (TO), Italy
}

   \date{Received ...; accepted ...}

\abstract{
 Moving groups in the solar neighborhood are ensembles of co-moving stars, likely originating due to forces from spiral arms, the Galactic bar, or external perturbations. Their co-movement with young clusters indicates recent star formation within these moving groups, but a lack of precise three-dimensional position and velocity measurements has obscured this connection. Using backward orbit integrations of 509 clusters within ~1 kpc—based on Gaia DR3 and supplemented with APOGEE-2 and GALAH DR3 radial velocities—we trace their evolution over the past 100 Myr. We find that most clusters separate into three spatial groups that each trace one of the Pleiades, Coma Berenices, and Sirius moving groups. The same trend is not seen for the Hyades moving group. The young clusters of the Alpha Persei, Messier 6, and Collinder 135 families of clusters, previously found to have formed in three massive star-forming complexes, co-move with either the Pleiades (Alpha Persei and Messier 6) or Coma Berenices (Collinder 135). Our results provide a sharper view of how large-scale Galactic dynamics have shaped recent, nearby star formation.
}

   \keywords{Galaxy: structure -- Galaxy: solar neighborhood -- Galaxy: stellar content}

   \maketitle

\section{Introduction}

    Building on the foundational work of \cite{proctor_ii_1870} and Kapteyn (1905), Olin J. Eggen \citep{eggen_concentrations_1983, eggen_star_1996} extensively cataloged co-moving stellar associations—termed stellar streams—with those observable in the solar neighborhood referred to as classical ``moving groups’’ (hereafter simply ``moving groups’’).The main moving groups in the solar neighborhood include the Pleiades group (also known as the “Local Association”), the Hyades group, the Coma Berenices group, and the Sirius group (Ursa Major). Each was initially proposed to originate from a dissolving open cluster or massive association \citep{eggen_concentrations_1983}, a concept referred to hereafter as the ``birth scenario''.
    
    Studies using precise astrometric data from missions like \textit{Hipparcos} and \textit{Gaia}, have confirmed the presence and internal structure of moving groups \citep{chen_identification_1997, dehnen_distribution_1998, gaia_collaboration_gaia_2018}, but also challenged their origin as coeval birth associations due to stellar age heterogeneity. For instance, \cite{famaey_age_2008} showed that most stars in the Pleiades moving group do not share the cluster’s age, ruling out a common origin. Numerous observational and numerical studies now suggest that classical moving groups are primarily kinematic structures spanning broad age ranges (several Gyr), likely shaped by internal disk dynamics—such as spiral arms, the Galactic bar, and their resonances \citep{quillen_effect_2005, quillen_spiral_2018, barros_exploring_2020}—and/or external perturbations from satellite galaxies \citep{minchev_low-velocity_2010, craig_dynamically_2021}

    A central question concerning moving groups is how and why they include recently formed stars among their members. Recently, \cite{quillen_birth_2020} linked recent star formation within 150 parsecs to the  dynamics of the moving groups —suggesting that large-scale perturbations, such as those from spiral arms, play a role in shaping the kinematics and spatial distributions of young stars. Gaia DR2-based studies also suggest clusters may trace moving groups \citep{soubiran_open_2018, tarricq_3d_2021}. Progress now hinges on extending this analysis to a larger, more precise sample—newly enabled by recent star cluster catalogs with precise 3D velocities out to 1 kpc \citep{hunt_improving_2023}.  
    
    In this \textit{Letter}, we leverage a sample of star clusters with precise 3D velocity measurements extending to distances of 1 kpc from the Sun. We trace their orbital histories and find clear associations with the prominent moving groups in the solar neighborhood. We first describe our data set and selection methods, then detail our orbital integration and clustering techniques, and finally discuss the implications of linking cluster formation and recent star formation history to the large-scale dynamical evolution of the Galactic disk and interstellar medium.

    \begin{figure*}[ht!]
        \centering
        \includegraphics[width=1\linewidth]{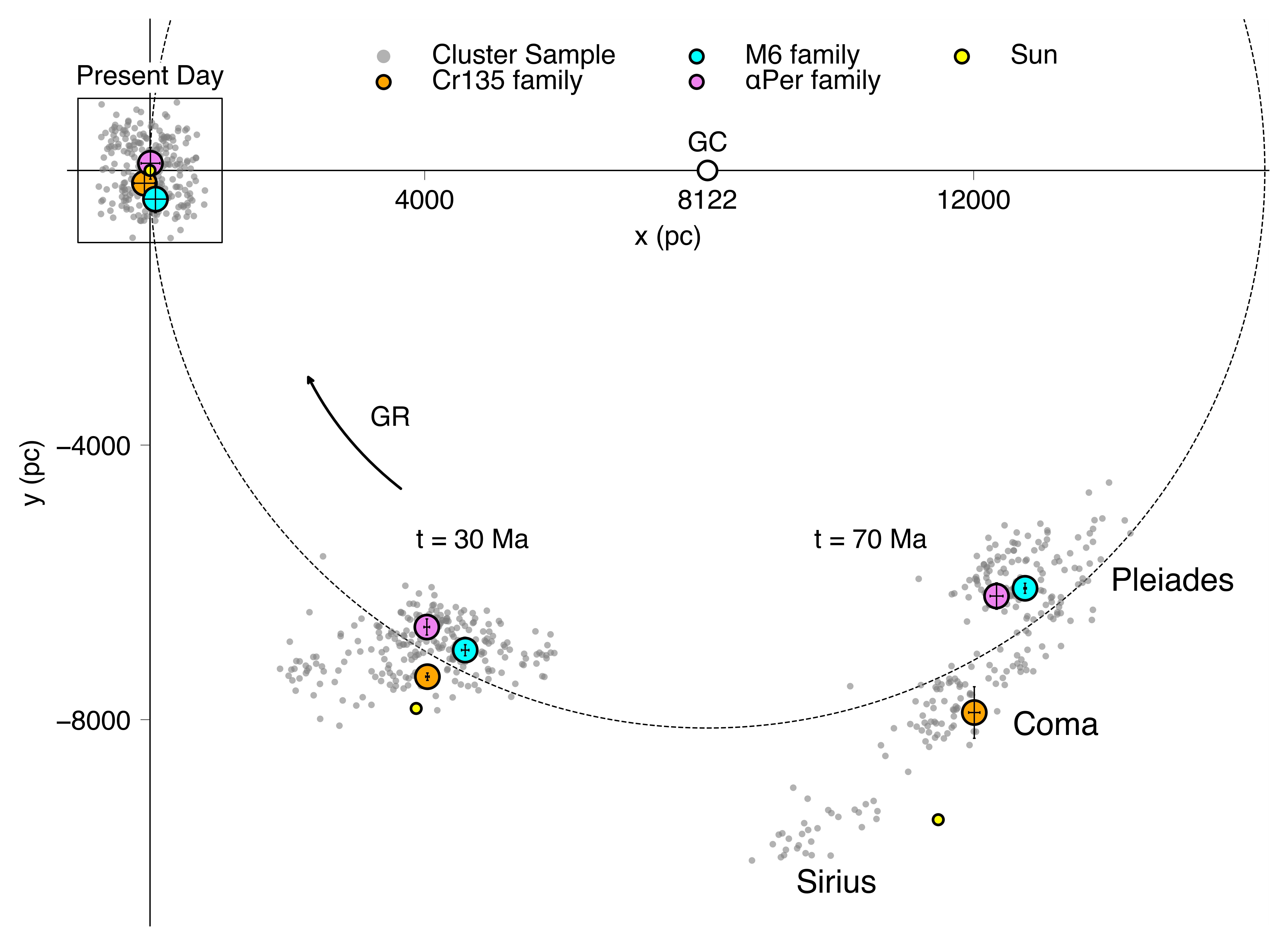}
        \caption{A Galactic bird’s-eye (XY) view of the clusters’ orbits over time, showing over one half of of a circle (black, dashed line) of radius of $R_{\text{Sun}}=8122$ pc extending from the Galactic Center to the Sun. The location of the Galactic Center and the direction of Galactic rotation are indicated. Individual clusters with ages between 70 and 150 Myr are represented as gray dots and are shown at three different times: the present (upper left), 30, and 70 million years ago (from left to right). The median location of each cluster family is displayed at the three time steps, with each family color-coded and labeled in the legend. The black bars for a given family indicate the standard deviation of its members’ positions at each time step. Average positional uncertainties of the older clusters are too small to show, but at $t = -70$ Myr they reach ~60 pc, slightly smaller than the grey point sizes. The three-dimensional interactive version of this figure has a time-slider showing the cluster positions at intermediate time steps with each frame centered on the location of the LSR: \url{https://cswigg.github.io/cam_website/interactive_figures/swiggum+25_interactive_figure1.html}.}
        \label{fig:Figure1}
    \end{figure*}
    
\section{Data}
    We use the \textit{Gaia} DR3-constructed catalog from \cite{hunt_improving_2023}, which uncovered 7167 Galactic star clusters using the HDBSCAN algorithm \citep{mcinnes_hdbscan_2017} . Although many nearby young clusters are no longer gravitationally bound \citep{hunt_improving_2024}, we refer to them as ‘clusters’ since they are each composed of stars that share a common origin and remain at least kinematically clustered. 

    \cite{swiggum_most_2024} (S24) cross-matched cluster members from \cite{hunt_improving_2023} with the APOGEE-2 DR17 \citep{abdurrouf_seventeenth_2022} and GALAH DR3 \citep{buder_galah_2021} surveys to supplement \textit{Gaia} DR3 radial velocities. They selected a high-quality subsample of nearby clusters (heliocentric Cartesian coordinates: $-1000<x,y<1000$ pc and $-300<z<300$ pc) with reliable three-dimensional velocities $(U,V,W)$, yielding 764 clusters before age cuts. From these, they retained 254 young clusters (age < 70 Myr) and added 27 Young Local Associations (YLAs)\footnote{Young Local Associations (YLAs), often called moving groups, are sparse, co-moving, co-eval stellar groups that differ from the more populous, non-co-eval classical moving groups.} from \cite{gagne_banyan_2018}. S24 found that most (155) clusters group into three families—Collinder 135 (39 clusters), Messier 6 (34 clusters), and Alpha Persei (82 clusters)—each converging to a former, massive star-forming complex. We use the positions, velocities, and ages of these 155 clusters, with S24’s computed errors, in our analysis.
    
    To investigate whether the cluster families show co-movement with older clusters, we select an additional subset of clusters from the high quality sample of 764 clusters mentioned above, but with ages older than the cluster families ($> 70$ Myr). This subset contains 509 clusters, with a median age of 153 Myr and $(U, V, W)$ velocity uncertainties of $(0.79, 0.81, 0.22)\mathrm{~km~s^{-1}}$.

\section{Methods and Results}
    \begin{figure*}[ht!]
        \centering
        \includegraphics[width=\linewidth]{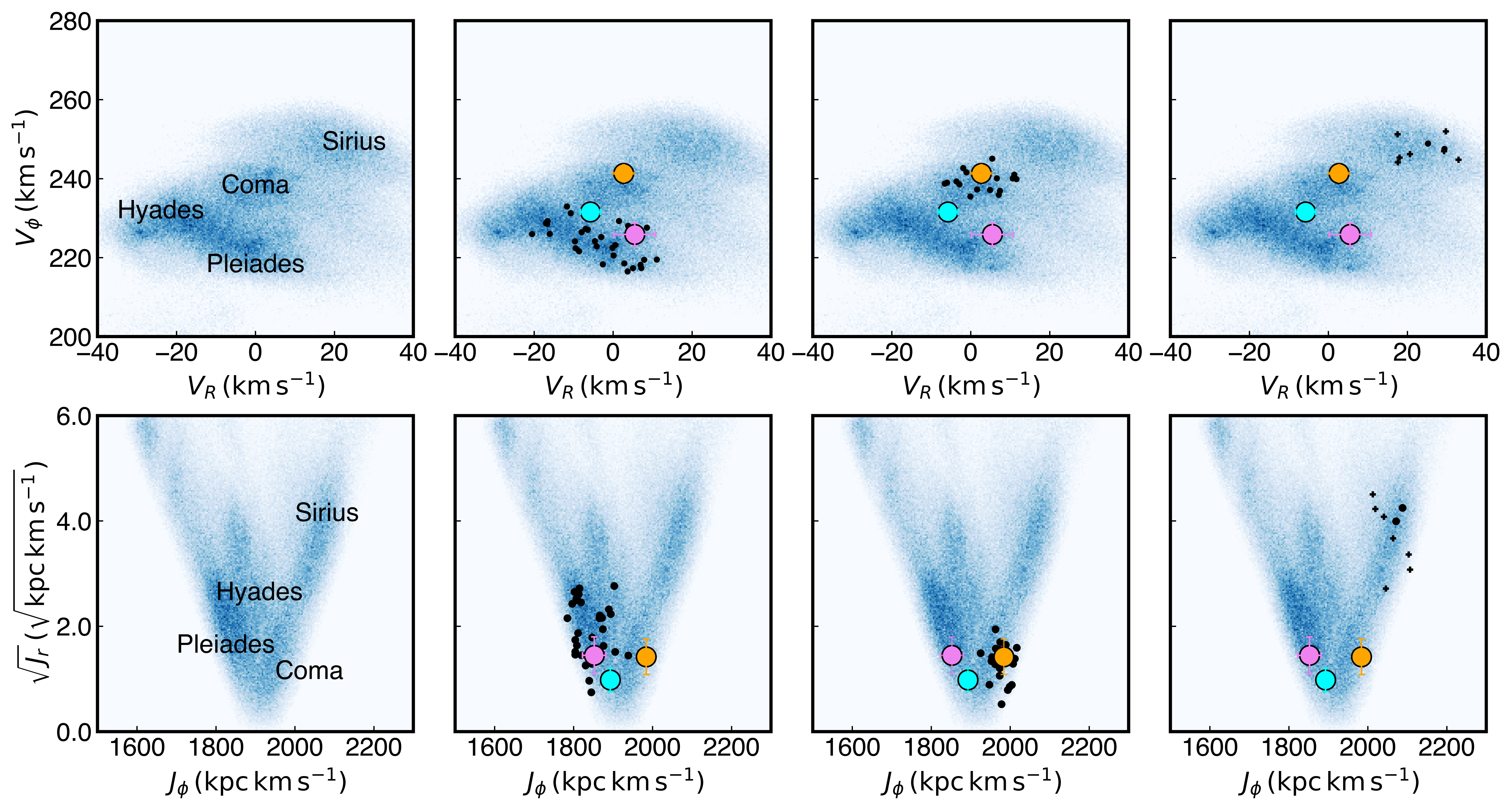}
        \caption{\textbf{Top row}: A 2D histogram ($V_R - V_\phi$; white-to-blue colormap) shows the distribution of roughly 1.8 million stars within 300 pc, revealing arch-like overdensities corresponding to the Pleiades, Coma Berenices, and Sirius moving groups (labeled in the first panel). The second to fourth panels overlay black dots for clusters older than 70 Myr, each showing one HDBSCAN-identified group. Panels two and three include only clusters within 300 pc to match the stellar volume; panel four shows a group extending to 500 pc, with clusters beyond 300 pc marked as crosses. Each group aligns with a distinct moving group. Bulk velocities of the $\alpha$Per (violet), M6 (cyan), and Cr135 (orange) families are shown with standard deviation error bars. \textbf{Bottom row}: Same as the top row, but showing $J_\phi - \sqrt{J_R}$ distributions instead.}
        
        \label{fig:Figure2}
    \end{figure*}

    \subsection{Computing orbital trace-backs}\label{orbits}
        Following \citet{swiggum_most_2024}, we use \verb|galpy| \citep{bovy_galpy_2015} and the \verb|MWPotential2014| model—comprising disk, bulge, and halo components—to integrate cluster orbits backward in time. We adopt a circular velocity of $236 ~\mathrm{km~s}^{-1}$, solar Galactocentric radius $R_\odot = 8122 , \mathrm{pc}$ \citep{reid_trigonometric_2019}, and solar height $z = 20.8 , \mathrm{pc}$ \citep{bennett_vertical_2019}. Cluster orbits are initialized with present-day heliocentric Cartesian positions $(x, y, z)$ and velocities $(U, V, W)$, corrected to the Local Standard of Rest using solar peculiar motion $(U_\odot, V_\odot, W_\odot) = (11.1, 12.24, 7.25), \mathrm{km,s}^{-1}$ \citep{schonrich_local_2010}.

        We integrate orbits 100 Myr into the past with 0.1 Myr time steps—balancing reliability against growing uncertainties from measurement errors (e.g., 1 km s$^{-1}$ translates to ~100 pc at 100 Ma) and unmodeled effects like spiral arm interactions. For the Cr135, M6, and $\alpha$Per cluster families, we compute bulk orbits by taking the median $(x, y, z)$ positions of member clusters at each time step. As these families formed after 70 Myr ago, earlier epochs approximate the motion of their natal molecular clouds. We include error bars showing spatial dispersion among member clusters. 
        
        \begin{table}
          \caption{Moving group cluster statistics: column 2 shows the number of clusters, column 3 the total number of stars of the clusters, and column 4 the mean cluster age with 16th–84th percentile spread.}
            \label{table:1}
            \centering
            \begin{tabular}{c c c c}
            \hline\hline
            Name & N & N$_{\text{stars}}$ & Age (Myr) \\
                \hline
            Pleiades & 154 & 35935 & $184^{+72}_{-97}$ \\
            Coma & 100 & 18932 & $189^{+56}_{-85}$ \\
            Sirius & 44 & 8258 & $229^{+140}_{-108}$ \\
            Not grouped & 211 & 46076 & $313^{+223}_{-208}$ \\
            \hline
            \end{tabular}
        \end{table}  

    \subsection{Examining cluster orbits}
        Figure \ref{fig:Figure1} shows the past Galactic $XY$ positions of the older cluster sample (gray points) with ages between 70 and 150 Myr, highlighting their locations at the present day, 30 Myr, and 70 Myr in the past.  The median cluster family locations are overplotted at each timestep. Starting around $t = 70$ Myr, the clusters separate into three distinct overdensities. Although our full sample of ‘older’ clusters includes those older than 150 Myr, we focus on the 70–150 Myr range in this figure, where the overdensities are most clearly, and curiously, apparent; older clusters tend to blur these features. Notably, the young families appear within two of the older overdensities in Figure \ref{fig:Figure1}. From $t = 70$ Myr ago to the present, the $\alpha$Per and M6 families traveled radially outward (away from the Galactic center) with a group of older clusters, while the Cr135 family moved slightly inward with a different group. This radial motion was calculated by \cite{swiggum_most_2024} as $\Delta R \approx 400$ pc for the $\alpha$Per and M6 families, and $\Delta R \approx -250$ pc for the Cr135 family over a similar 60 Myr time-span.

        A key distinction between the young cluster families from \cite{swiggum_most_2024} and the older cluster overdensities seen here is that the former converge near their formation times, implying a common origin in the same gas complex. This is evident in the \href{https://cswigg.github.io/cam_website/interactive_figures/swiggum+25_interactive_figure1.html}{interactive version} of Figure \ref{fig:Figure1}, where the individual clusters of the families are shown and this convergence can be seen. The older overdensities at $t = 70$ Myr (Figure \ref{fig:Figure1}), by contrast, consist of co-moving clusters that do not necessarily converge or share a common formation origin, but rather a common dynamical origin. Still, since most clusters are younger than a Galactic orbit (~200 Myr), and the overdensities are most prominent in the 70–150 Myr range, it’s plausible that some clusters in this older sample originated from the same natal region—suggesting they belong to older cluster families. Identifying older cluster families requires improved observational constraints and is limited by our knowledge of the Galactic potential \citep{arunima_their_2025}.

    \subsection{Identifying cluster groups with HDBSCAN}\label{sec:hdbscan}
        To robustly select cluster members of the three spatial overdensities (hereafter `groups') visible in Figure \ref{fig:Figure1}, we employ the method of \cite{swiggum_most_2024}. At each time step starting from $t = 50$ Myr ago (the time when the spatial groups first become visually apparent in the cluster orbits) to $t = 100$ Ma, we apply the \verb|HDBSCAN| clustering algorithm \citep{mcinnes_hdbscan_2017} to the clusters’ 3D positions. We find that setting \verb|min_cluster_size| to 30 and \verb|cluster_selection_method| to `leaf’ recovers these groups. At each time step, \verb|HDBSCAN| assigns each star cluster as either a member of an identified group or as `noise'. To determine the final group membership, we compute the most frequently assigned label for each cluster across all time steps and assign this label as its final group. This process is repeated 100 times, with each iteration randomly sampling from the position and velocity uncertainties of the star clusters. Using these parameters, we consistently recover three distinct groups, with no additional groups detected. Changing the value of \verb|min_cluster_size| can affect the number of recovered groups, but since our goal is to robustly and reproducibly recover the three groups visible by eye, we do not explore this parameter extensively.
        
    	We report the statistics of these groups in Table (\ref{table:1}). In the following section, we show that these groups become visually apparent (\ref{sec:kinematic_plane}) in the past, as they each trace one of the Pleiades, Coma Berenices, and Sirius moving groups, consisting of 154, 100, and 44 clusters, respectively, from the initial sample of 509 clusters. 
        
    \subsection{Examining velocities and actions}\label{sec:kinematic_plane}

        The \textit{Hipparcos} and \textit{Gaia} missions have revealed kinematic substructure in the solar neighborhood, especially in radial versus tangential velocities ($V_R$–$V_\phi$) \citep{dehnen_local_1998, gaia_collaboration_gaia_2018, lucchini_moving_2022, bernet_ridges_2022}. We query the \textit{Gaia} database for stars with measured radial velocities and high-quality parallaxes within 300 pc ($d \approx 1/\varpi$) of the Sun and convert their ICRS coordinates to a Galactocentric cylindrical frame using \verb|astropy.coordinates| \citep{astropy_collaboration_astropy_2022}. The same process is applied to our older cluster sample and cluster families, yielding positions $(R,\phi,z)$ and velocities $(V_R,V_\phi,V_z)$ (corrected for solar motion; see Section \ref{orbits}).

        The distribution of azimuthal ($J_\phi$) and radial ($J_R$) actions—where $J_\phi$ is angular momentum and $J_R$ traces orbital eccentricity \citep{trick_galactic_2019}—offers another way to identify moving groups \citep{coronado_pearls_2022, furnkranz_age_2024}. In our axisymmetric potential (\verb|MWPotential2014|), actions computed during orbital integrations are displayed in the second row of Figure \ref{fig:Figure2}.
        
        In the top row of Figure \ref{fig:Figure2}, a 2D histogram of the $V_R$–$V_\phi$ plane for nearby stars shows three arch-shaped overdensities. These arches correspond to the Hyades, Pleiades, Coma Berenices, and Sirius (Ursa Major) moving groups, although only three cluster groups (overlaid in subsequent panels) align with the arches—namely, the Pleiades, Coma, and Sirius groups. The bottom row displays the $J_\phi$–$J_R$ distribution, where moving groups appear as diagonal overdensities. Clusters in the second and third columns are restricted to 300 pc (matching the star sample), while those in the fourth column extend to 500 pc due to their scarcity within 300 pc.
        
        Each cluster group aligns with one of the three velocity arches, as labeled in Figure \ref{fig:Figure1}. The median $V_R$–$V_\phi$ velocities of the cluster families are shown, with the $\alpha$Per and M6 families overlapping the Pleiades arch—albeit offset toward higher circular velocities. Although the Hyades cluster (Melotte 25) appears in the star sample, it is not recovered as part of any cluster group; its region is weak in the clusters’ distribution, though some Hyades contribution is suggested in the action distributions. Consequently, we refer to these clusters as the Pleiades moving group, while acknowledging possible Hyades ``contamination''. The Cr135 family appears associated with the Coma moving group.

\section{Discussion and Summary}

    To summarize, our results link recent star formation to the complex velocity structure driven by large-scale Galactic dynamics. The three cluster families from \cite{swiggum_most_2024} likely formed in large star-forming complexes within ~1 kpc over the past 70 Myr, shaping and driving energy and momentum into the interstellar medium via stellar feedback \citep[e.g.][]{soler_kinetic_2025}. In this \textit{Letter}, we find that the Alpha Persei and Messier 6 families align with the Pleiades moving group, whereas the Collinder 135 family aligns with the Coma Berenices group. Notably, the clusters do not closely follow the Hyades moving group—nor is the Hyades cluster recovered in our \verb|HDBSCAN| analysis.

    Moving groups have long been known to host young stars and clusters \citep[e.g.,][]{eggen_concentrations_1983, soubiran_open_2018}. Using our ability to trace cluster motions backward in time (Figure \ref{fig:Figure1}), we link young clusters to these groups. The three cluster families likely formed in massive star-forming complexes whose inherited motions, shaped by Galactic-scale perturbations, align with older clusters influenced by the same forces. These older clusters form three past over-densities—corresponding to the Pleiades, Coma Berenices, and Sirius groups (Figure \ref{fig:Figure2}). Although Eggen’s co-eval formation model is insufficient, we argue it remains relevant: young clusters forming within giant gas complexes—initially comoving with older stars of the moving group—drift apart after expelling their natal gas via stellar feedback. This is in agreement with \cite{liang_moving_2024}, who find evidence of enhanced star formation in the groups.
    
    The Pleiades moving group (or “Local Association”) includes young stars spanning 5–120 Myr—too broad for a single coeval origin \citep{fernandez_kinematic_2008}. Sco-Cen, IC 2602, the Alpha Persei cluster (Melotte 20), and the Pleiades cluster have long been recognized as notable members \citep{eggen_concentrations_1983}. Backward integration of the Alpha Persei family by \citet{swiggum_most_2024} reveals that its youngest members—Sco-Cen and Taurus—converge with the 60 Myr-old Alpha Persei cluster, suggesting a sequential formation scenario within a single gas complex. This process may have been driven by feedback-induced bubbles from earlier generations of clusters, which collapse and trigger new star formation \citep{elmegreen_sequential_1977}. In contrast, older co-moving clusters like the Pleiades likely did not form in this complex but share similar kinematics due to their moving group membership. As discussed in Section 3, older cluster families might exist among the 70-150 Myr old clusters, but shared origins over 100 Myr ago are harder to confirm due to increasing uncertainties with time.
       
    The kinematic coherence between the Alpha Persei and M6 cluster families - together with the older clusters of the Pleiades moving group - suggests a common formation mechanism, possibly influenced by the dynamics of the spiral arm. Considering a simple model, gas entering a spiral arm is slowed by gravitational torques and pressure forces, leading to density enhancements that trigger star formation. Newly formed stars inherit the gas velocity and eventually settle in stable orbits with reduced radial velocities relative to the Galactic center, drifting outward over 60–120~Myr while maintaining similar azimuthal speeds. The radial velocities for the Alpha Persei family, the M6 family, and older Pleiades group clusters are typically between 5-6 km/s. According to the latest data of the Milky Way having a bar with a pattern speed of $\sim 39$km/s/kpc \citep{gaia_collaboration_gaia_2023_drimmel} the 4:1 resonance of a spiral pattern of 18-20 km/s/kpc is likely to be located at approximately 7.5 kpc \citep{quillen_effect_2005}, inside the solar radius. This resonance, likely associated with the Carina arm, could become a natural site for gas compression and cluster formation.
    
    Spiral shock scenarios have previously been considered for many of the YLAs—most of which are members of the Alpha Persei family—by \cite{quillen_birth_2020}, and also for the Sco-Cen complex, also part of the Alpha Persei family, by \cite{fernandez_kinematic_2008}. Recent findings that the solar neighborhood is currently close to the co-rotation radius of the spiral pattern might also explain why the moving groups persist in ages up to ages of Gyr, much older than the clusters considered in our work \citep{barros_exploring_2020}.
 
    Alternatively, resonance trapping could maintain stellar orbits in stable configurations near specific resonances.
    Given the Milky Way’s bar pattern speed, its Outer Lindblad Resonance (OLR) is located far beyond the solar neighborhood \citep{portail_chemodynamical_2017, donghia_trojans_2020, lucchini_milky_2023}. However, the 4:1 resonance associated with the Carina arm at 7.5 kpc, in addition to determining where star formation is enhanced, might also play a role in trapping stellar systems. At a 4:1 resonance, the condition is given by
    \[
     \Omega_\phi = 4\Omega_r 
    \]
    where \(\Omega_r\) is the radial oscillation frequency and \(\Omega_\phi\) is the azimuthal frequency. Moreover, if the orbital frequencies follow power-law relations, one expects a correlation in the orbital actions such that
    \[
    J_r \propto J_\phi^{-\beta/\alpha},
    \]
    with \(\alpha\) and \(\beta\) describing the Galactic potential. However, Figure \ref{fig:Figure2} shows that the orbital actions for these clusters do not show such clear correlations between $J_r$ and $J_\phi$, suggesting that resonance trapping is unlikely to explain their coherent kinematics. Instead, transient spiral arm features, which briefly compress gas and then locally dissolve \citep{donghia_self-perpetuating_2013} provide a more natural explanation for the similarities observed.
    
    \textit{Gaia} has ushered in an era where we can, in 3D and across time, visualize the spatial and temporal evolution of star formation across a vast volume of the solar neighborhood. Our backward orbit integrations reveal a clear link between recent star formation and Galactic dynamics—likely driven by spiral arms. Future work will explore the age distribution of clusters in detail, especially as the 70–150 Myr population most distinctly outlines the moving groups. With Gaia DR4 offering improved astrometry and radial velocities, it will be key to advancing our understanding of the Galaxy’s structure and star formation history.

\begin{acknowledgements}
We thank Robert Benjamin and Sebastian Ratzenböck for helpful discussions. Co-funded by the European Union (ERC, ISM-FLOW, 101055318). This work has made use of data from the European Space Agency (ESA) mission Gaia \url{https://www.cosmos.esa.int/gaia}, processed by the Gaia Data Processing and Analysis Consortium (DPAC, \url{https://www.cosmos.esa.int/web/gaia/dpac/consortium}). Funding for the DPAC has been provided by national institutions, in particular the institutions participating in the Gaia Multilateral Agreement.
\end{acknowledgements}

\bibliographystyle{aa}
\bibliography{references, references-2}
\end{document}